\documentclass[twocolumn,showpacs,preprintnumbers,amsmath,amssymb]{revtex4}
\usepackage{graphicx}
\usepackage{dcolumn}
\usepackage{bm}
\usepackage{graphics}
\usepackage{psfrag}
\usepackage{epsfig}
\usepackage{amssymb}

\begin{document}

\title{Mutual Information as a Tool for Identifying Phase Transitions in Dynamical Complex Systems with Limited Data}

\author{R. T. Wicks}

\author{S. C. Chapman}

\author{R. O. Dendy}
\altaffiliation[Also at ]{UKAEA Culham Division, Culham Science Centre, Abingdon, Oxfordshire, OX14 3DB, UK.}
	 \affiliation{Centre for Fusion, Space and Astrophysics, University of Warwick, Coventry, CV4 7AL, UK.}

\date{\today}

\begin{abstract}

We use a well known model (T. Vicsek \textit{et al.} Phys Rev Lett \textbf{15}, 1226 (1995)) for flocking to test mutual information as a tool for detecting order-disorder transitions, in particular when observations of the system are limited. We show that mutual information is a sensitive indicator of the phase transition location, in terms of the natural dimensionless parameters of the system which we have identified. When only a few particles are tracked, and when only a subset of the positional and velocity components are available, mutual information provides a better measure of the phase transition location than the susceptibility of the data.

\end{abstract}

\maketitle

\section{Introduction}

Order-disorder transitions are often found in complex systems. They have been identified in physical systems such as Bose-Einstein condensates and ferromagnets and in biological, chemical and financial systems. Phase transitions are found, for example, in the behaviour of bacteria \cite{Czirok1}, locusts \cite{Buhl}, voting games and utilisation of resource in markets \cite{Savit}. These systems have in common the property that there is competition between fluctuations driving the system towards disorder, and inter-element interactions driving the system towards order. Insight into such systems can be gained using simple models. Although the dynamics of individual elements are difficult to predict, one can identify macroscopic parameters that characterise the behaviour of the system. These can be approached through dimensional analysis, e.g. Buckingham's $\Pi$ theorem \cite{Longair}.
\paragraph*{}
A generic challenge in real world measurements of physical, chemical, biological or economic systems is that they yield datasets that are, in essence, sparse. Single elements such as tracer particles in turbulent flow, tagged birds or dolphins in a group or a constituent of a financial index may, or may not, adequately sample the full underlying system behaviour. In consequence, the behaviour of a finite number of individual elements may, or may not, provide a proxy for the behaviour of the entire system. If the system behaviour is known to exhibit a phase transition, the question arises how this can best be captured from analysis of the dynamics of individual elements. Previously, for example, both mutual information (MI) \cite{Shannon} \cite{Cellucci} \cite{Kraskov} \cite{TKMarchRecPlot} and susceptibility have been shown to be sensitive to the phase transition in the Ising spin model of ferromagnetism \cite{Matsuda}. MI can also extract correlation, or dependence, between causally linked but spatiotemporally separated observed parameters: for example, between in-situ plasma measurements in the solar wind, and the ionospheric response detected by ground based measurements on Earth \cite{TKMarch}; and within the brains of Alzheimer's disease patients \cite{EEG}.
\paragraph*{}
Here we compare the use of MI and susceptibility to quantify the location of the phase transition in the dimensionless parameter space of the Vicsek \textit{et al.} model \cite{Vicsek}. There are numerous statistical methods for analysing systems with many degrees of freedom, dating back to work by Helmholtz and Boltzmann in the 19th century, see for example \cite{Coveney} \cite{Penrose} and references therein. The microscopic behaviour of the Vicsek system, described below, does not conserve energy or momentum and this precludes a microscopic understanding of energy, momentum or any related quantities in the system. The driving characteristic of the system is entropy: this is augmented by the random forcing of the particles, and removed by their mutual interactions, which create correlation. With this in mind, we examine mutual information as a natural choice for capturing the entropy flow in the system, and characterising the system state with respect to its order parameters.
\paragraph*{}
We find that when full knowledge of the system is available, that is when all the particles are tracked, the susceptibility is an accurate method for estimating the position of the phase transition in the Vicsek model. However, if the data is limited to a sample of just a few particles out of a large number, or a subset of the complete data, this method is less accurate. 
\paragraph*{}
We show that the mutual information of only a few particles, or of limited data from the whole system, can successfully locate the phase transition in dimensionless parameter space. For example we find that the MI of a timeseries of components of particle position or velocity is sufficient. We thus show that MI can provide a practical method to detect order-disorder transitions when only a few particles, or elements, of the system are observed.

\section{The Vicsek Model}
\begin{figure}[t]
	\centering
\includegraphics{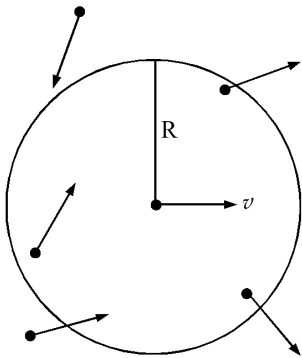}
\caption{Multiple particles interact if within a radius $R$ of each other. Each of the $N_R$ particles within $R$ (here $N_R = 4$) contributes its angle of propagation to the average $\langle \theta^{N_R}_{n} \rangle$, which is assigned to the particle at the centre of R.} \label{fig:interact}
\end{figure}

In 1995 Vicsek \textit{et al.} \cite{Vicsek} introduced the self propelled particle model in which particles have a constant speed $|\underline{v}| = v_{0} $ and a varying direction of motion $\theta$. In the discrete time interval $\delta t = t_{n+1} - t_{n}$ an isolated particle increments its vector position $\underline{x}_{n} \rightarrow \underline{x}_{n+1}$ by moving with constant speed $v_{0}$ in a direction $\theta_{n}$ which is in turn incremented at each timestep. In the model, particles interact when they are within distance $R$ of each other, such that the direction of their motion tends to become oriented with that of their neighbours. This interaction is implemented at each step, as shown in Fig \ref{fig:interact}, by replacing the particle's direction of motion $\theta_n$ by the average of those particles $N_{R}$ within distance $R$, so that $\theta _{n+1} = \langle \theta^{N_{R}}_{n} \rangle$ with a random angle $\delta \theta_{n}$ also added. The random fluctuation $\delta \theta_{n}$ is an independent identically distributed angle in the range $-\eta \leq \delta \theta_{n} \leq \eta$, where $\eta$ characterises the strength of the noise for the system. Thus for the $i^{th}$ particle in the system, after $n$ timesteps:
\begin{eqnarray}
\underline{x}^{i}_{n+1} & = & \underline{x}^{i}_{n} + \underline{v}_{n}^{i}\, \delta t \label{eq:stepx} \\ 
\theta^{i} _{n+1} & = & \langle \theta^{N_{R}}_{n} \rangle + \delta \theta^{i}_{n} \label{eq:stept} \\
\underline{v}^{i}_{n} & = & v_0 \left(\cos \theta^i_n \underline{\hat{x}} + \sin \theta^i_n \underline{\hat{y}} \right) \label{eq:velstep}
\end{eqnarray}
Here, direction is defined by the angle from the x axis $(\underline{\hat{x}})$, and $\eta$ is such that $\eta = \overline{\eta} \delta t$, that is normalised to the time step $\delta t$.

\begin{figure}[h!t]
\centering

\resizebox{3.6in}{!}{\includegraphics{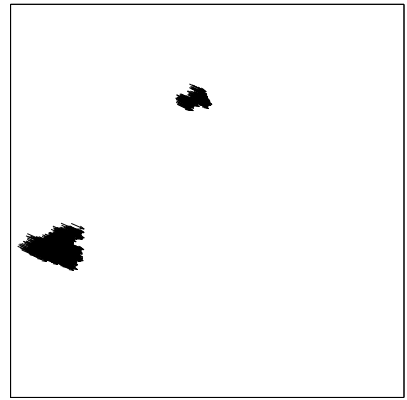}}

\resizebox{3.6in}{!}{\includegraphics{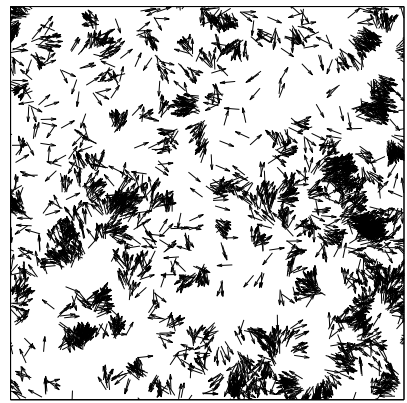}}

\resizebox{3.6in}{!}{\includegraphics{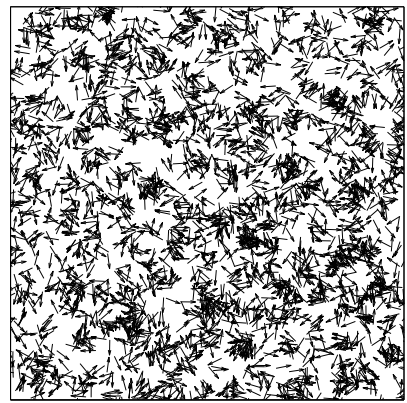}}

\caption{The effect of increasing noise on a typical Vicsek system from ordered dynamics (top: $\eta = 0$) to disordered dynamics (bottom: $\eta = 4\pi/5$), and in the vicinity of the phase transition (middle: $\eta = 2\pi/5$). Particle velocity vectors are plotted as arrows at the position of each particle in the x-y plane. The system has parameters $N = 3000$, $|v| = 0.15$, $R = 0.5$. This corresponds to $\Pi_{2} = 0.3$, $\Pi_{3} = 0.94$ and $\Pi_{1} = \eta$, see equations \ref{eq:pi1}-\ref{eq:pi3}.} \label{fig:BirdPics}
\end{figure}

\begin{figure}[h!t]
    \includegraphics{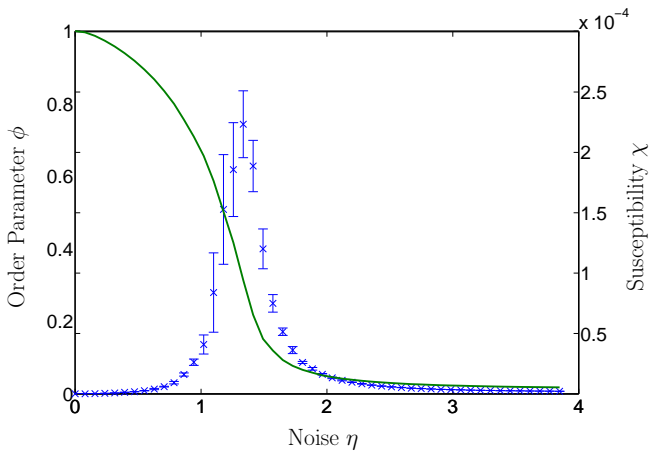}
    \caption{(Colour online) An example of a typical Vicsek system. The order parameter $\phi$ (line) is maximum for zero noise and falls to a constant small value at high noise. The susceptibility $\chi$ (crosses) peaks at the critical point $\eta_c \approx 1.33$ for the system. The system parameters are: $\Pi_2=0.3$ and $\Pi_3=0.94$, with $N = 3000$.} \label{fig:Ordersus}
\end{figure}
\paragraph*{}
There are two limiting cases for the system dynamics: disorder, where each particle executes a random walk; and order, where all particles move together with the same velocity. Figure \ref{fig:BirdPics} shows snapshots of the system dynamics for $\eta = 0$, $\eta = 2\pi/5 \simeq 1$ and $\eta = 4\pi/5 > 1$. We see that $\eta \ll 1$ is highly ordered and $\eta \gg 1$ is highly disordered, and around $\eta \simeq 1$ there is a phase transition \cite{Gregoire1}. As with other critical systems it is possible to define an order parameter $\phi$ and a susceptibility $\chi$ \cite{Vicsek} \cite{Gregoire2} \cite{Czirok2} \cite{Czirok3}. For the Vicsek model, the magnitude of the average velocity of all the particles in the system provides a macroscopic order parameter and the variance of this speed is the susceptibility:
\begin{eqnarray}
\phi &=& \frac{1}{Nv_0} \left| \sum^{N}_{i=1} \underline{v}^{i} \right|\\ \label{eq:orderparam}
\chi &=& \sigma^2(\phi) = \frac{1}{N} \left( \langle \phi^2 \rangle - \langle \phi \rangle^2 \right) \label{eq:suscept}
\end{eqnarray}
Here $N$ denotes the total number of particles in an implementation of the model of Vicsek \textit{et al.}.
\paragraph*{}
We plot $\phi$ and $\chi$ as a function of $\eta$ in Fig \ref{fig:Ordersus}. In the thermodynamic limit $(N \rightarrow \infty, l \rightarrow \infty)$ where $l$ is the system size, the susceptibility would tend to infinity at the critical noise $\eta_c$, where the phase transition occurs. In a finite sized realisation of system, the susceptibility has a sharp but finite maximum at the critical noise at which the phase transition occurs. Finite size effects make the peak location uncertain, but it is still possible to obtain an estimate of the critical noise $\eta_c$.
\paragraph*{}
The system can be analysed using Buckingham's $\Pi$ theorem \cite{Longair}, and three independent dimensionless quantities can be formed that characterise its behaviour. The first of these ($ \Pi_{1}$) is the amplitude of the normalised noise $\eta$, the second ($\Pi_{2}$) is the ratio of the distance travelled in one timestep $v_0 \delta t$ to the interaction radius $R$, and the third ($\Pi_{3}$) is the average number of particles contained within a circle of one interaction radius $R$:
\begin{eqnarray}
\Pi_{1} & = & \eta = \overline{\eta} \delta t \label{eq:pi1} \\ 
\Pi_{2} & = & v_0\delta t/R \label{eq:pi2} \\ 
\Pi_{3} & = & \pi R^{2}\rho \label{eq:pi3} 
\end{eqnarray}
These three parameters determine the behaviour of the system in the thermodynamic limit $(N \rightarrow \infty$, $l \rightarrow \infty$, $R$ and $\rho$ finite$)$ where $\rho$ denotes the number density of particles over the whole system.
\paragraph*{}
The system size $l$ affects the number of interactions that occur. If $l$ is finite and the system is periodic as here, the finite system size increases the chance of two randomly chosen particles interacting, compared to the limit of infinite $l$. The system only approaches the thermodynamic limit when the finite interaction radius $R \ll l$. Conversely, for example, if the interaction radius is half the diagonal size of the system, then all the particles interact with each other at any given moment. This implies a fourth parameter reflecting the finite size of any computer based realisation of this model:
\begin{eqnarray}
\Pi_{4} & = & R/l \label{eq:pi4}
\end{eqnarray}
In the thermodynamic limit we have $N \rightarrow \infty, l \rightarrow \infty$, whilst $R$ and $\rho = N/l^2$ are finite, so that $\Pi_4 \rightarrow 0$ and $\Pi_{1-3}$ are finite and specify the system. % The results obtained by changing these parameters are shown in the Results section below. 
\section{System Phase Space}
\begin{figure}[t]
    %\centering
    \includegraphics{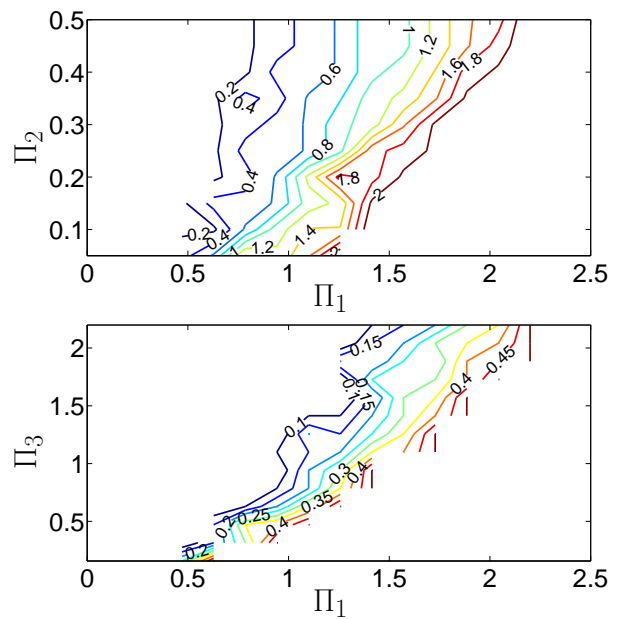}
    \caption{(Colour online) Phase transition diagram contours for the Vicsek model around $\Pi_3 = 1$. Top panel: the effect of changing $\Pi_3$, from $\Pi_3 = 0.2$ (dark blue contours, left hand side) to $\Pi_3 = 2.0$ (dark red contours, right hand side) in steps of 0.2, on the position of the phase transition in the $\Pi_1, \Pi_2$ plane. Bottom panel: the effect of changing $\Pi_2$, from $\Pi_2 = 0.05$ (dark blue, left hand side) to $\Pi_2 = 0.5$ (dark red, right hand side) in steps of 0.05, on the position of the phase transition in the $\Pi_1, \Pi_3$ plane.} \label{fig:PhasePlot}
\end{figure}

For given values of $\Pi_2$ and $\Pi_3$, we run simulations of the Vicsek system for a range of values of $\Pi_1$ to determine the value $\Pi_1 = \Pi_1^c$ at which the susceptibility $\chi$ peaks and thus the phase transition occurs. By repeating this operation for a set of parameter values of $\Pi_2$ and $\Pi_3$, we obtain the full set of coordinates at which the phase transition is located for a region of the phase space around $\Pi_3 = 1$. We show this graphically in Fig \ref{fig:PhasePlot} where we plot contours of $\Pi_3^c(\Pi_1, \Pi_2)$ in the upper panel, and $\Pi_2^c(\Pi_1,\Pi_3)$, in the lower panel. These plots confirm that there is a smooth, well defined surface of $\Pi_1^c, \Pi_2^c, \Pi_3^c$; they can be used to inform the choice of $\Pi_1, \Pi_2$ and $\Pi_3$ for the next section.
\paragraph*{}
In relation to recent work by Nagy \textit{et al.} \cite{Nagy}, we see from Fig. \ref{fig:PhasePlot} that the speed $v_0$ of the particles, which is contained within $\Pi_2$, has a characteristic effect above $v_0 \approx 0.3$. The bottom panel of Fig \ref{fig:PhasePlot} shows that, for constant $\Pi_3$, the phase transition becomes hard to detect and the contours start to break up as $\Pi_2$ is increased above approximately $0.3$. This is a complementary demonstration of the statement made in \cite{Nagy} that the phase transition becomes first order in the high velocity $(v_0 \geq 0.3)$ regime.

\section{Mutual Information}

Mutual information (MI) quantifies the information content shared by two signals $A$ and $B$. For discrete signals we can write the MI as

\begin{equation}
I(A,B) = \sum_{i,j}^m P(a_i, b_j) \log_2 \left( \frac{P(a_i, b_j)}{P(a_i)P(b_j)} \right) \label{eq:MI1}
\end{equation}

Here the signal $A$ has been partitioned into an alphabet (a library of possible values the signal can take) $A = \{a_1, \ldots, a_i, \ldots a_m\}$ where $a_1$ and $a_m$ are the extrema of $A$ found in all data considered. The discretized signal takes value $a_i$ with probability $P(a_i)$ and similarly for $b_i$ we have $P(b_i)$, while $P(a_i, b_j)$ is the joint probability of $a_i$ and $b_j$ occurring together. The chosen base of the logarithm defines the units in which the mutual information is measured. Normally base two is used, so that the mutual information is measured in bits. If we define the entropy of a signal as
\begin{eqnarray}
H(A) & = & -\sum_{i}^m P(a_i)\log_2(P(a_i)) \label{eq:MI2}
\end{eqnarray}
then MI can be written as a combination of entropies \cite{Shannon}

\begin{eqnarray}
I(A,B) & = & H(A) + H(B) - H(A,B) \label{eq:MI3}
\end{eqnarray}

The calculation of the entropies needed to form the MI is not trivial, as there is some freedom in the method of discretization of the signals and in the method used to estimate the probabilities $P(a_i)$, $P(b_j)$ and $P(a_i,b_j)$. There are many different methods currently used, summarised and compared by Cellucci \textit{et al.} \cite{Cellucci} and Kraskov \textit{et al.} \cite{Kraskov}.
\paragraph*{}
MI has been used in the analysis of the two dimensional Ising model by Matsuda \textit{et al.} \cite{Matsuda}. Importantly the critical temperature for the Ising model is picked out precisely by the peak in the mutual information of the whole system. This peak survives the coarse graining of the system very well, which raises the possibility of mutual information being useful in the study of other complex systems.

\section{Identifying the Phase Transition}

\subsection{Full System Mutual Information}

In the 2D Vicsek system there are three variables for each of the $N$ particles: their positions $(x^i, y^i)$ and the orientation of their velocities $\theta^i$, giving three signals $X$, $Y$ and $\Theta$ each containing $N$ measurements at every time step. The simplest discretization of these signals $x^i$, $y^i$, $\theta^i$ is to cover the range of the signals with equally spaced bins, so for position coordinate $X$ we have $m$ bins $X_i$ with width $\delta X$. Then if $n$ particles are in the range $(X_i - X_i+\delta X)$ we have probabilities:
\begin{eqnarray}
P(X_{i}) = \frac{n(X_{i})} {N \delta X} \\
\sum_{i} P(X_{i}) = 1
\end{eqnarray}
The single and joint probabilities $P(Y_{j})$, $P(\Theta_{k})$, $P(X_{i}, \Theta_{k})$ and $P(Y_{j}, \Theta_{k})$ are calculated in a similar manner.
\paragraph*{}
The key factor governing the accuracy with which MI is measured is to optimise the size of the bins used in the above procedure. If the bins are too large then resolution is lost, and the exact level of small scale structure and clustering cannot be identified. If the bins are too small then at high noise the probability of finding a particle at a given point does not become smoothed over the whole system because individual particles can be resolved, giving $P(x_i, y_j) \neq P(x_i)P(y_j)$ even though the system is in a well mixed random state.
\paragraph*{}
There is no ideal bin structure yet determined for this method of MI calculation \cite{Cellucci} \cite{Kraskov}. The Vicsek model has two natural length scales, $R$ the interaction radius and $l$ the box size, so that a good length scale to choose for discretization, when a snapshot of the whole system is being used, is the interaction radius $R$. Thus all our mutual information calculations made on the whole system use a bin size of $2R$, the diameter of the circle of interaction; the bins are therefore squares of size $4R$ in the $(x,y)$ plane. When $\theta$ is discretized the same number of bins are used as for $x$ or $y$ because there is no natural size for bins in $\theta$.
\paragraph*{}
Given full knowledge of $x_i, y_i, \theta_i$ for all $N$ particles in the system over a large number of timesteps, several different calculations of mutual information can be made. We find that the most accurate form of mutual information for the whole system is that calculated between $x$ or $y$ position and $\theta$. Thus we perform the following calculation at each time step $n$ once the system has reached a stable state:
\begin{eqnarray}
I(X,\Theta) &=& \sum_{i,j} P(X_i, \Theta_j) \log_2 \left( \frac{P(X_i, \Theta_j)}{P(X_i)P(\Theta_j)} \right)\;\;\; \label{eq:SpecMI1}\\
I(Y,\Theta) &=& \sum_{i,j} P(Y_i, \Theta_j) \log_2 \left( \frac{P(Y_i, \Theta_j)}{P(Y_i)P(\Theta_j)} \right)\;\;\; \label{eq:SpecMI2}\\
I &=& \frac{I(X,\Theta) + I(Y,\Theta)}{2} \label{eq:SpecMI3}
\end{eqnarray}
and average over all timesteps for which MI is measured.

\begin{figure}[t]
    \centering
    \includegraphics{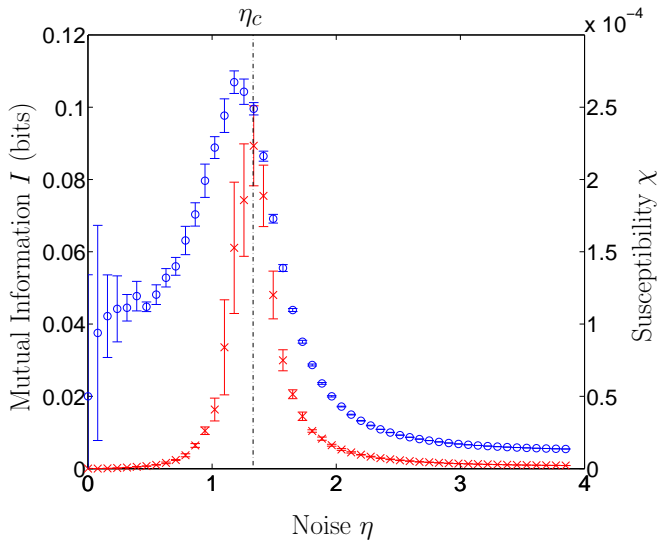}
    \caption{(Colour online) The mutual information, $I$ (circles) defined by Eq. (\ref{eq:SpecMI3}) peaks at approximately the same point as the susceptibility, $\chi$ (crosses) defined by Eq. (\ref{eq:suscept}) the critical noise $\eta_c \approx 1.33$ is marked. The system parameters are $\Pi_2=0.15$ and $\Pi_3=0.98$, with $N = 3000$ particles. Error bars on the susceptibility are largest around $\eta_c$, unlike those on mutual information.} \label{fig:MIsus}
\end{figure}

We compare the MI as calculated using the above method and the susceptibility as a function of normalised noise $\eta$ in Fig \ref{fig:MIsus}. At large $\eta$ the MI falls to zero as $X, Y$ and $\Theta$ tend to uncorrelated noise (see also \cite{Matsuda}). We would also expect the MI to fall to zero at sufficiently low $\eta$ as the system becomes ordered and this behaviour is also seen within the errors. The errors on our measurements of MI are calculated from the standard deviation of measurements of MI calculated over 50 simulations at each noise $\eta$. The error on the susceptibility is calculated in the same manner.
\paragraph*{}
The error bars become larger at low $\eta$ because the mutual information includes the signatures of spatial clustering as well as velocity clustering in the measurement. Thus at low $\eta$, when extended clusters form, the mutual information will give a higher value for the more spatially extended axis of the cluster and a lower value for the less extended axis of the cluster. This implies that the shape and orientation of the (usually single) large cluster formed at low noise influences the mutual information. Different measurements of MI thus arise for each implementation of the model, giving rise to the error seen at low $\eta$. This could be corrected by using other approaches to computing MI, for example recurrence plotting \cite{TKMarchRecPlot} \cite{TKMarch}, or a different distribution of bins; these are more computationally intensive, however.
\paragraph*{}
When estimated as the standard deviation over 50 repeated runs of the simulation, the error is found to be considerably larger, as a fraction of the overall measurement, for the susceptibility than for the MI. This is because the susceptibility is simply an average fluctuation over all the velocity vectors of the system; whereas the MI also directly reflects the level of spatial `clumpiness' (that is, spatial correlation) of the particles. The detailed spatial distribution varies from one simulation to the next, but at fixed $\Pi_{1-4}$ the degree of `clumpiness' does not. Mutual information is able to quantify clustering (correlation in space as well as velocity) in a simple dynamical complex system, in a manner that identifies the order-disorder phase transition.

\subsection{Mutual Information from Limited Data}

Observations of many real world systems typically provide only a subset of the full system information, which here comprises the positions and velocities of all $N$ interacting particles. We now consider results from the Vicsek model using only very limited amounts of data. The mutual information and susceptibility are now calculated on a $\tau = 5000$ step timeseries of positional and velocity data for $n = 10$ particles out of the $N = 3000$ simulated. To optimise both methods, the data for each particle timeseries is cut into $S$ sections, labelled $s = 1, \ldots ,S$ of length $N_s = \tau/S$ steps. This gives us $nS$ pseudo-systems, relying on the assumption that one particle over $N_s$ steps is equivalent to $N_s$ particles at one step. This is a reasonable assumption to make for the Vicsek model as it is ergodic while $\eta$ remains constant.
\paragraph*{}
To calculate the susceptibility, we need to estimate the variance of the average velocity of each of these $nS$ pseudo-systems. We therefore cut each section $s$ into $S'$ further subsections $s'$, calculate the average velocity $\phi^i_{s'}$ of these subsections and find their variance, giving $\chi^i_{s}$ the pseudo-system susceptibility. This is done for each pseudo-system individually to give $\chi^i_s$, and averaged over all $nS$ pseudo-systems to give $\overline{\chi}$, the average variance of the average velocity for all pseudo-systems:

\begin{eqnarray}
\phi^i_{s'} & = & \frac{SS'}{\tau v_0}\left|\sum^{\tau/SS'}_{k=1} \underline{v}^i_k \right| \label{eq:PartOrderParam} \\
\chi^i_{s} & = & \frac{1}{S'}\left( \langle \phi_{s'}^2 \rangle - \langle \phi_{s'} \rangle^2 \right) \label{eq:PartChi} \\
\overline{\chi} & = & \frac{1}{nS} \sum_{s=1}^S \sum_{i=1}^n \chi^i_s \label{eq:ChiTot}
\end{eqnarray}

\begin{figure}[t]
    \centering
    \includegraphics{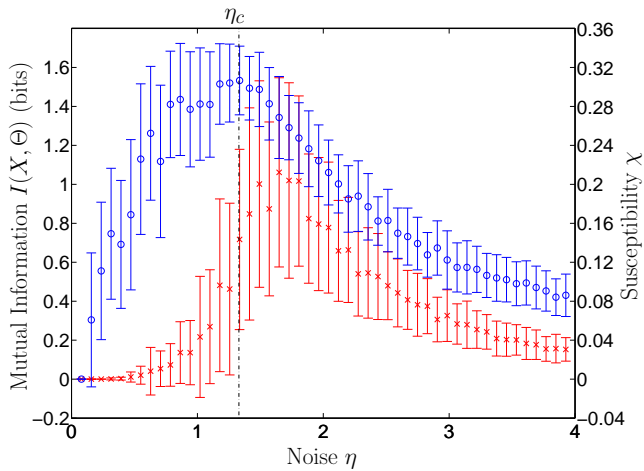}
    \caption{(Colour online) The mutual information $I$ (circles) calculated using timeseries from only ten particles for 5000 time steps, with $S = 10$, compared to the average susceptibility $\overline{\chi}$ (crosses) for the same data using $S' = 10$ subsections to calculate $\overline{\chi}$ and with the critical noise $\eta_c \approx 1.33$ marked. System parameters are $\Pi_2=0.3$ and $\Pi_3=0.94$, with N $N = 3000$ particles.} \label{fig:MIsus2}
\end{figure}

The result is shown in Fig \ref{fig:MIsus2} where we also plot the mutual information $I(X,\Theta)$ from Eq. (\ref{eq:SpecMI1}), but now as $nS$ timeseries; the parameters used are $n = 10$, $S = 10$, $S' = 10$. The error bars are calculated as the standard deviation of the 100 measurements made using the different pseudo-systems of length $\tau/S = 500$ timesteps. These values for $n$, $S$ and $S'$ are chosen so as to limit the data in a realistic way. $n = 10$ is a suitably small subset of the $N = 3000$ particles. $S = 10$ cuts the data into segments sufficiently long (500 timesteps) to be treated independently. $S' = 10$ is chosen so that each section $s'$ is still long enough (50 timesteps) to make as good an estimate of the average velocities $\phi^i_{s'}$ as possible, but allows enough of these measurements to be made to reduce the error on the measurement of $\chi^i_{s}$.

\begin{figure}[t]
    \centering
    \includegraphics{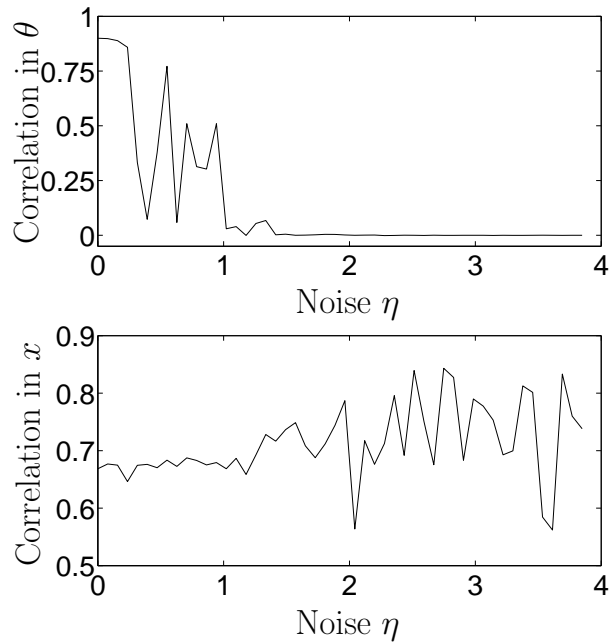}
    \caption{The cross correlation between a randomly chosen particle and nine others, calculated using a timeseries with 5000 steps. The top panel shows the average cross correlation between $\theta^1$ and $\theta^k, 2 \leq k \leq 10$. The bottom panel shows the average cross correlation between $x^1$ and $x^k, 2 \leq k \leq 10$. System parameters are $\Pi_2=0.3$ and $\Pi_3=0.94$, with $N = 3000$ particles.} \label{fig:CrossCorr}
\end{figure}

\paragraph*{}
The system is identical to that shown in Fig \ref{fig:MIsus} and the phase transition is at the same noise, $\eta_c \approx 1.33$. Near their respective peaks, the error in the mutual information remains smaller than that in the susceptibility and so MI better identifies the peak. The peak in the susceptibility no longer coincides with $\eta_c$ and is shifted to the higher noise side of the phase transition. This occurs because the susceptibility is now measured on too small a sample of data: only 50 angles $\theta^i_t$ are averaged to find each subsection velocity $\phi^i_{s'}$. Such a small ensemble average results in a large deviation in the average velocity from the expected value.
\paragraph*{}
For comparison with a linear measure, we calculate the cross correlation for our ten trajectories. We choose one of the particles at random and compute its cross correlation with each of the remaining nine. The average of these is plotted in Fig \ref{fig:CrossCorr}. The average cross correlation between angles $\theta^1$ and $\theta^k, 2 \leq k \leq 10$ in the top panel shows strong correlation at low noise, as expected. This cross correlation declines as noise increases, but not smoothly, because the correlation depends on the exact dynamics of the particles considered. Angular cross correlation reaches zero around the phase transition, but does not provide an accurate location for the critical noise. In the bottom panel of Fig \ref{fig:CrossCorr} the cross correlation between $x^1$ and $x^k, 2 \leq k \leq 10$ provides no reliable indication of the position of the phase transition. The cross correlation does become more variable on the higher noise side of the graph but this effect cannot be used to accurately find the critical noise $\eta_c$.
\paragraph*{}
The value of using mutual information can be seen when the available data is restricted still further. Let us consider signals derived from one component of the particle trajectory only, equivalent to a line of sight measurement. We then have one of the position coordinates $x^i_k$, and the instantaneous $x$ component of the velocity, $\Delta x^i_k = v_0 cos(\theta^i_k)$. The susceptibility is calculated as in Eqs (\ref{eq:PartOrderParam}) - (\ref{eq:ChiTot}), but using the average one dimensional velocities $\Delta x^i_k$: 
\begin{eqnarray}
\phi^i_{s'} & = & \frac{SS'}{\tau v_0}\left|\sum^{\tau/SS'}_{k=1} \Delta x^i_k \right| \label{eq:DeltaOrderParam} \\
\chi^i_{s} & = & \frac{1}{S'}\left( \langle \phi_{s'}^2 \rangle - \langle \phi_{s'} \rangle^2 \right) \label{eq:DeltaChi} \\
\overline{\chi} & = & \frac{1}{nS} \sum_{s=1}^S \sum_{i=1}^n \chi^i_s \label{eq:DeltaChiTot}
\end{eqnarray}
The mutual information is calculated for each section of the $x$ only (and later $y$ only) components of the timeseries for each particle using a suitable binning:
\begin{equation}
I(X,\Delta X) = \sum_{i,j} P(X_i, \Delta X_j) \log_2 \left( \frac{P(X_i, \Delta X_j)}{P(X_i)P(\Delta X_j)} \right) \label{eq:DeltaMIX}
\end{equation}
\begin{equation}
I(Y,\Delta Y) = \sum_{i,j} P(Y_i, \Delta Y_j) \log_2 \left( \frac{P(Y_i, \Delta Y_j)}{P(Y_i)P(\Delta Y_j)} \right) \label{eq:DeltaMIY}
\end{equation}

\begin{figure}[t]
\centering
   \includegraphics{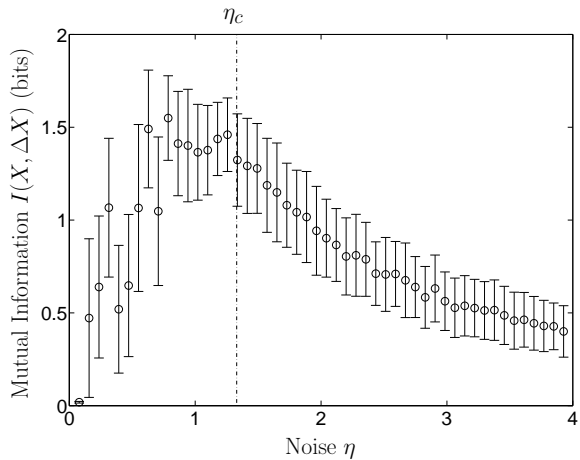}
    \caption{The mutual information $I(X,\Delta X)$ (circles) calculated using a timeseries from only ten particles for 5000 time steps, with $S = 10$ and the critical noise $\eta_c \approx 1.33$ marked. System parameters are $\Pi_2=0.3$ and $\Pi_3=0.94$, with $N = 3000$ particles.} \label{fig:MI10binXXYY}
\end{figure}
\begin{figure}[t]
\centering
    \includegraphics{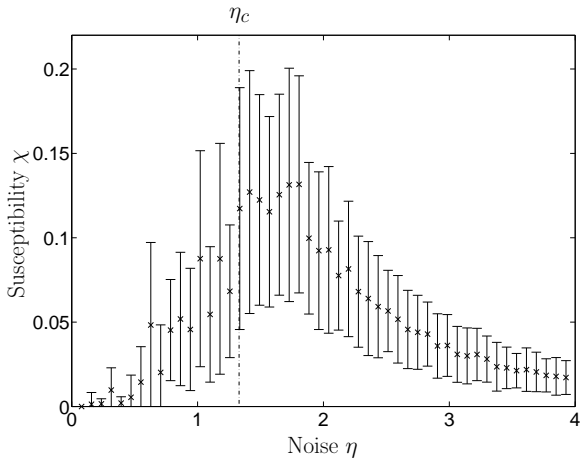}
    \caption{The susceptibility $\chi$ (crosses) calculated using a timeseries of one dimensional data $\{X, \Delta X\}$ from only ten particles for 5000 time steps, with $S = 10$ and the critical noise $\eta_c \approx 1.33$ marked. System parameters are $\Pi_2=0.3$ and $\Pi_3=0.94$, with $N = 3000$ particles.} \label{fig:Sus10binXX}
\end{figure}

\paragraph*{}
Figure \ref{fig:MI10binXXYY} shows the mutual information calculated from the data in this manner with $S = 10$. The peak in the mutual information is at approximately the correct value of $\eta$ ($\eta_c \approx 1.33$). Figure \ref{fig:Sus10binXX} shows for comparison the susceptibility calculated over the $X$ data as in equations (\ref{eq:DeltaOrderParam})-(\ref{eq:DeltaChiTot}). We see that although there is a peak, it no longer identifies $\eta \rightarrow \eta_c$ accurately. The peak is broader and has larger error bars than in Fig \ref{fig:MI10binXXYY}, giving a large uncertainty in identifying $\eta_c$.
\begin{figure}[t]
\centering
    \includegraphics{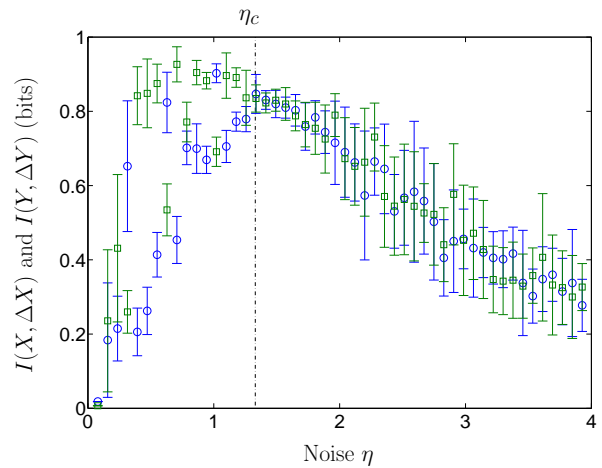}
    \caption{(Colour online) The mutual information $I(X,\Delta X)$ (circles) and $I(Y,\Delta Y)$ (squares) calculated using a timeseries from only ten particles for 5000 steps, with $S = 1$ and the critical noise $\eta_c \approx 1.33$ marked. System parameters are $\Pi_2=0.3$ and $\Pi_3=0.94$, with $N = 3000$ particles.} \label{fig:MI2binXXYY}
\end{figure}
\paragraph*{}
The peak in Fig \ref{fig:MI10binXXYY} is shifted to the low noise side of the phase transition and shows some scatter. This can be understood by looking at the same data using a different value of the interval $S$. In Fig \ref{fig:MI2binXXYY} we show the same data analysed using $S = 1$, that is we consider one timeseries of length 5000 time steps for each of ten particles and obtain MI averaged over these ten. We plot $I(X,\Delta X)$ (circles) and $I(Y,\Delta Y)$ (squares). The measurements overlap within errors on the high noise side of the phase transition but separate into two distinct branches, containing both $I(X,\Delta X)$ and $I(Y,\Delta Y)$, on the low noise side.

\begin{figure}[t]
\centering
    \includegraphics{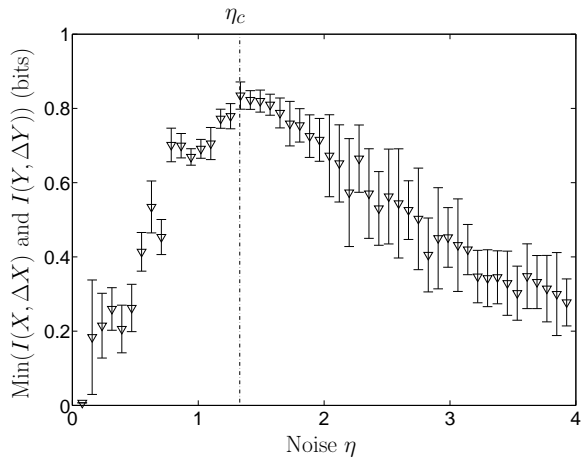}
    \caption{The minimum results from mutual information measurements $I(X,\Delta X)$ and $I(Y,\Delta Y)$ calculated using a timeseries from only ten particles for 5000 steps, with $S = 1$ and the critical noise $\eta_c \approx 1.33$ marked. System parameters are $\Pi_2=0.3$ and $\Pi_3=0.94$, with $N = 3000$ particles.} \label{fig:MIminXXYY}
\end{figure}

\paragraph*{}
One potential source of this behaviour is that, as the system becomes ordered at $\eta < \eta_c$, the particles clump together. This implies that the particles together take on a preferred direction of motion; in addition, a clump may be elongated in a particular spatial direction. The effectiveness of MI will then depend on whether our single component (line of sight) data is aligned along, or perpendicular to, these key directions. Mutual information measured in terms of coordinates aligned with the preferred direction of motion is increased by the dispersion of particle positions and velocities, whereas MI measured in term of perpendicular coordinates is decreased because the positional and velocity dispersion are smaller in this direction. Anomalously high MI measurements result from making measurements along the preferred direction of motion; large relative velocities lead to anomalously high peaks on the low noise side of the phase transition, making it appear to be shifted towards $\eta = 0$.
\paragraph*{}
Finally in Fig \ref{fig:MIminXXYY} we plot the minimum of $I(X,\Delta X)$ and $I(Y,\Delta Y)$ for each value of $\eta$ from Fig \ref{fig:MI2binXXYY} and see that a clear peak emerges at $\eta = \eta_c$, where the error bars are smallest. This outcome obviates the difficulty that arises if we only allow knowledge of $\{X, \Delta X\}$ for example, when it would be necessary to exclude high measurements of MI at low noise, as discussed above.

\section{Conclusions}

The Vicsek model \cite{Vicsek} is used here to test the potential of measurements of order and clustering that exploit mutual information in dynamic complex systems. We find that when complete knowledge of the system is available, the mutual information has a smaller error than the susceptibility (Fig \ref{fig:MIsus}). Using Buckingham's $\Pi$ theorem the set of dimensionless parameters that capture the phase space of the Vicsek model have been presented as a complete set for the first time.
\paragraph*{}
When data is limited to observations of only ten particles out of 3000, the error in the mutual information remains comparatively small, and the mutual information thus provides a better measurement than susceptibility of the position of the order-disorder phase transition (Fig \ref{fig:MIsus2}). When data is limited still further, such that only one line of sight component of the particle motion is available, the mutual information measurement remains sensitive enough to identify the critical noise of the phase transition, while the susceptibility does not (Figs \ref{fig:MI10binXXYY} - \ref{fig:MIminXXYY}).
\paragraph*{}
In this case the mutual information also provides an indication of the preferred axial direction of clumped particle motion at low noise. Anomalously high mutual information estimates in this ordered phase indicate that the particles measured are mostly moving along the dimension being measured; low estimates indicate that the particles are moving perpendicular. This is remarkable given that susceptibility does not contain this information, and that the MI is a probabilistic measurement.
\paragraph*{}
Real world data are often in the form of the final data studied here; a limited sample from a much larger set, measured in fewer dimensions than those of the original system: for example, line of sight measurements of wind speed measured by an anemometer at a weather station, or satellite measurements of the solar wind. It has been shown here that mutual information can provide an effective measure of the onset of order, and may provide a viable technique for real world data with its inherent constraints.\\

\section*{Acknowledgements}
This work was supported in part by the EPSRC. RW acknowledges a PPARC CASE PhD studentship in association with UKAEA. The authors would like to thank Khurom Kiyani for valuable discussions and the Centre for Scientific Computing at the University of Warwick for providing the computing facilities.

 \end{document}